# Spectroscopic Signatures of Interlayer Coupling in Janus MoSSe/MoS$_2$ Heterostructures

Kunyan Zhang[1], Yunfan Guo[2,*], Daniel T. Larson[3], Ziyan Zhu[3], Shiang Fang[4], Efthimios Kaxiras[3,5], Jing Kong[2,*], Shengxi Huang[1,*]

[1]Department of Electrical Engineering, The Pennsylvania State University, University Park, Pennsylvania 16802, United States

[2]Department of Electrical Engineering and Computer Science, Massachusetts Institute of Technology, Cambridge, Massachusetts 02139, United States

[3]Department of Physics, Harvard University, Cambridge, Massachusetts 02138, United States

[4]Department of Physics and Astronomy, Center for Materials Theory, Rutgers University, Piscataway, New Jersey 08854, United States

[5]John A. Paulson School of Engineering and Applied Sciences, Harvard University, Cambridge, Massachusetts 02138, United States

*Corresponding author

**Abstract**

The interlayer coupling in van der Waals heterostructures governs a variety of optical and electronic properties. The intrinsic dipole moment of Janus transition metal dichalcogenides (TMDs) offers a simple and versatile approach to tune the interlayer interactions. In this work, we demonstrate how the van der Waals interlayer coupling and charge transfer of Janus MoSSe/MoS$_2$ heterobilayers can be tuned by the twist angle and interface composition. Specifically, the Janus heterostructures with a sulfur/sulfur (S/S) interface display stronger interlayer coupling than the heterostructures with a selenium/sulfur (Se/S) interface as shown by the low-frequency Raman modes. The differences in interlayer interactions are explained by the interlayer distance computed by density-functional theory (DFT). More intriguingly, the built-in electric field contributed by the charge density redistribution and interlayer coupling also play important roles in the interfacial charge transfer. Namely, the S/S and Se/S interfaces exhibit different levels of PL quenching of MoS$_2$ A exciton, suggesting the enhanced and reduced charge transfer at the S/S and Se/S interface, respectively. Our work demonstrates how the asymmetry of Janus TMDs can be used to tailor the interfacial interactions in van der Waals heterostructures.

# INTRODUCTION

Janus transition metal dichalcogenides (TMDs) have attracted tremendous interest due to their mirror asymmetry-induced properties.[1-7] In monolayer Janus TMDs, the transition metal atoms are sandwiched between two different chalcogen layers, whose difference in electronegativity naturally leads to a built-in electric field in the direction perpendicular to the basal plane. This built-in electric field provides an additional degree of freedom with which to tune the van der Waals interaction between adjacent 2D layers. Our previous work has demonstrated that the out-of-plane dipole moment of Janus TMDs is capable of enhancing the van der Waals interlayer coupling by as much as 13% when compared to the corresponding TMD homobilayers.[8] Other works also investigated Janus TMD heterostructures with high-symmetry stackings by direct synthesis.[8-10] However, the full potential of Janus TMDs to tune the van der Waals interfacial coupling awaits further exploration.

The attempt to manipulate interlayer interactions by changing the twist angle between crystal axes has been triggered by the breakthroughs in magic-angle graphene[11, 12] and twisted TMD heterobilayers.[13-16] In the twisted van der Waals heterostructures, localized quantum states and enhanced electron correlation can arise as a result of the formation of a moiré superlattice. Besides the twist angle, the direction of the intrinsic dipole moment in Janus TMDs can have significant impacts on the fundamental physical properties. For example, the carrier separation and recombination can be manipulated by the intrinsic electric field of Janus TMDs.[17, 18] By stacking Janus TMDs to form homobilayer or -trilayers, a band offset is predicted to exist across the layers, which results in prolonged exciton lifetimes.[19] While the stacking configuration also plays an essential role in determining the excitonic behaviors.[19, 20] Moreover, it is predicted that the intrinsic dipole moment of Janus TMDs can be used to control the plasmon energy[21] and the Schottky barrier of graphene.[22, 23]

In this work, we employed the Janus TMD MoSSe to tune the van der Waals coupling and charge transfer between $MoS_2$ and MoSSe. The interfacial interactions were tuned through the twist angle between the MoSSe and $MoS_2$ atomic crystals and the direction of the intrinsic dipole moment, namely, whether the sulfur or selenium atoms of the Janus layer are adjacent to the top sulfur layer in $MoS_2$ (S/S and Se/S interfaces). For both types of interfaces, the interlayer shear mode is only evident for small twist angles, similar to the observation in conventional TMD bilayers. However, the S/S heterobilayer exhibits a stronger interlayer coupling than the Se/S heterobilayer as shown by the low-frequency Raman modes, which is consistent with the smaller interlayer distance calculated by density functional theory (DFT). Furthermore, the effects of interlayer distance and strain relaxation are revealed by the high-frequency Raman modes. Moreover, intralayer excitons, especially the A exciton of $MoS_2$, are significantly affected by the charge transfer process associated with the intrinsic dipole moment of MoSSe. The interfacial electron-hole separation is either promoted or inhibited by the built-in electric field at the S/S and Se/S interfaces, contributing to contrasting photoluminescence (PL) quenching. Our work demonstrates the possibility of using the asymmetry of Janus TMD to tailor the interfacial interactions in TMD heterostructures, providing an additional tool to design and fabricate Janus TMD-based devices for applications including optoelectronics, valleytronics, and spintronics.

## RESULTS AND DISCUSSION

### Stacking-Dependent Interlayer Breathing Modes

We fabricated Janus MoSSe/MoS$_2$ heterostructures with two types of interfaces including S/S and Se/S interfaces, in which the sulfur or selenium side of MoSSe is in contact with the bottom MoS$_2$ layer (Figure 1a). Monolayer MoSSe was prepared by the selenization of chemical vapor deposition (CVD) synthesized MoS$_2$,[2] and then transferred onto the as-grown MoS$_2$ by poly(methyl methacrylate) (PMMA) to form the S/S or Se/S interfaces (METHODS).[9] The bilayers were stacked with a relative angle $\theta$ between the crystal axes of MoSSe and MoS$_2$ with $\theta$ ranging from 0 to 60°. The twist angle is defined as the angle between the zigzag directions of MoS$_2$ and MoSSe as shown in Figure 1a. In our Raman analysis, the Raman frequency was averaged from the least square fitting of Stokes and anti-Stokes Raman spectra. (Typical anti-Stokes Raman spectra are shown in Figure S1.) As a result, a spectral resolution of 0.2 cm$^{-1}$ was achieved.

The low-frequency Raman spectra of the S/S heterobilayer exhibit clear interlayer shear and breathing modes that are related to interlayer atomic vibrations (Figure 1b). However, the low-frequency modes of MoSSe/MoS$_2$ heterobilayers made by transfer stacking are red-shifted compared to the heterobilayers fabricated by direct selenization with high-symmetry 2H and 3R stackings as shown in Figure 1b.[8] The redshift of the Raman frequencies of twisted heterobilayers is similar to that of twisted bilayer MoS$_2$ whose interlayer Raman mode frequencies downshift compared to exfoliated MoS$_2$ bilayers. This is explained by the stronger interlayer coupling in the exfoliated samples that have better interfacial quality.[24-27]

Another behavior similar to twisted bilayer MoS$_2$ is the dependence of the interlayer mode frequency on the twist angle. For example, the interlayer shear mode at around 20 cm$^{-1}$, arising from in-plane interlayer vibrations, is only evident for heterobilayers with twist angles close to 0 and 60° (Figure 1c, bottom panel), while the twisted MoSSe/MoS$_2$ heterostructures exhibit active interlayer breathing modes around 31-37 cm$^{-1}$ for all twist angles (Figure 1c, top panel). The breathing mode frequencies achieve the highest values for $\theta$ close to 0 and 60° and are red-shifted by 2-4 cm$^{-1}$ for intermediate twist angles between $\theta = 5$ and 55°. This can be understood by the changes in the local stacking geometry caused by the relative twist between the TMD layers. The MoSSe/MoS$_2$ heterobilayers have three high-symmetry stacking patterns at $\theta = 0°$ and three at $\theta = 60°$ that are related by in-plane translations (Figure S2).[24] For twisted samples with $\theta$ close to 0 and 60°, there remain large areas where the atoms are locally in a high-symmetry arrangement.[24, 28] However, for intermediate twist angles between $\theta = 5$ and 55°, the area of the local high-symmetry regions becomes smaller and even disappears, leading to reduced interlayer vibrational frequencies.[24, 28] The transition twist angles of 5 and 55° also coincide with the structural transition angle $\theta = 4°$ of twisted bilayer MoS$_2$ from the relaxed to the rigid regime as demonstrated by Quan *et al.*, at which the interlayer breathing mode has a frequency splitting up to 15 cm$^{-1}$.[28]

Figure 1d summarizes the twist-angle dependence of the measured breathing mode frequencies by grouping the measurements in bins of 10°. For intermediate twist angles between $\theta = 5$ and 55°, the MoSSe/MoS$_2$ heterostructure with the S/S interface exhibits breathing mode frequencies that reach a local maximum for $\theta$ close to 30°, as shown in Figure 1d, top panel. Similar to the S/S heterostructure, the Se/S heterostructure also shows higher breathing mode frequencies close to 30° (Figure 1d, bottom panel). This observation cannot be explained by the reduced area of high-symmetry regions described above, since the high-symmetry patterns disappear when $\theta$ exceeds 10°.[24] Even though several previous studies on twisted bilayer $MoS_2$ show similar twist-angle dependence, there is little discussion on the origin of this effect.[24, 28] In addition, the breathing mode frequency of the Se/S heterostructure is lower than that of the S/S heterostructure, especially for 5-15° and 45-55° (Figure 1d), which suggests a weaker interlayer interaction in the Se/S heterobilayer compared to the S/S heterobilayer. The origin of the different interlayer coupling of S/S and Se/S heterostructures is explained by the interlayer distance, which is discussed in detail later.

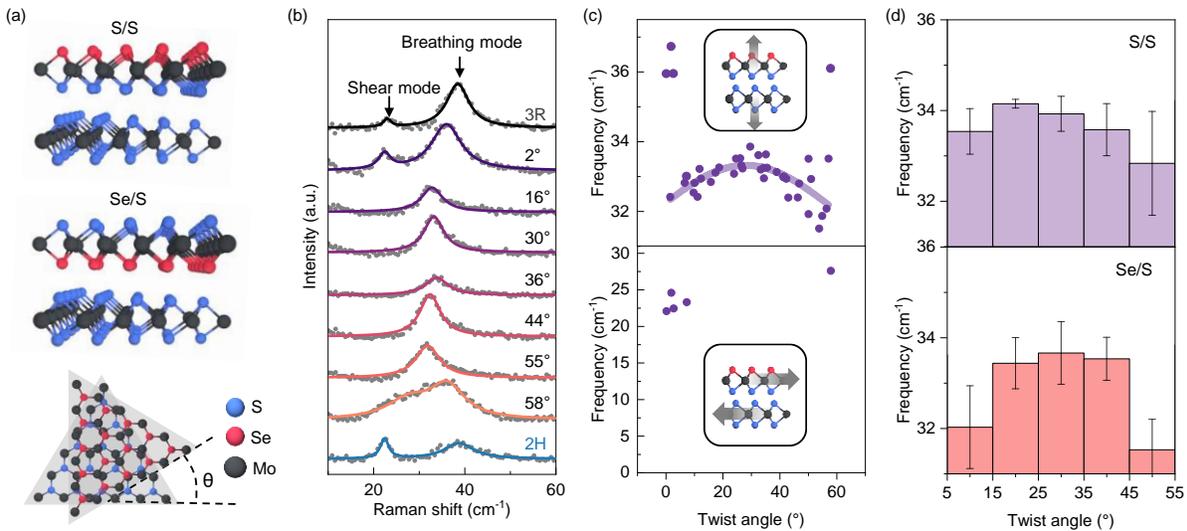

**Figure 1**. Low-frequency Raman modes of $MoSSe/MoS_2$ with S/S and Se/S interfaces. (a) Illustration of S/S and Se/S heterobilayers. The twist angle $\theta$ is defined as the angle between the zigzag directions of MoSSe and $MoS_2$. (b) Raman spectra of S/S heterobilayers produced by transfer stacking with different twist angles and fabricated by direct synthesis (2H and 3R). The gray dots are the measured data and the curves are the fitted peaks. (c) The experimental interlayer breathing mode (top panel) and shear mode (bottom panel) frequencies of the S/S heterobilayer for different twist angles. The arrows in the insets label the relative vibrations between layers for interlayer breathing and shear modes. The curve is a guide to the eye. (d) Summary of the interlayer breathing mode frequencies for S/S and Se/S heterobilayers at the intermediate twist angles for every 10°. The bins are the mean value, and the error bars are the standard deviation of the measured samples in each bin.

To understand the evolution of low-frequency modes as a function of the twist angle, we performed DFT calculations to determine the zone center phonon modes of the $MoSSe/MoS_2$ bilayers with S/S and Se/S interfaces, yielding the predicted Raman frequencies. We studied $MoSSe/MoS_2$ bilayers in commensurate supercells with twist angles of $\theta = 0°$, 13.2°, 21.8°, 27.8°, 32.2°, 38.2°, 46.8°, and 60°. For each twist angle, several distinct configurations are possible depending on the relative horizontal alignment of the two layers. We only considered the configurations with high-symmetry regions that most closely resemble the untwisted structures with low energy (Figure S2). Several of the angles allow similar but distinct structures. The specific atomic configurations used in the DFT calculations are shown in Figure S3. The calculated interlayer shear and breathing mode frequencies in Figures 2a-b are consistent with the experiments (Figures 1c-d). For $\theta = 0$ and 60°, the calculated shear mode frequencies are around 22-26 cm$^{-1}$, while the breathing mode frequencies are around 34-38 cm$^{-1}$, in agreement with the experimental values of 22-27 and 36-37 cm$^{-1}$. The absence of interlayer shear modes with $\theta$ away from 0 and 60° in our measurements is consistent with the predicted shear mode frequencies below 4 cm$^{-1}$ (Figure 2a), which is below the instrument detection limit. Compared to the interlayer breathing mode, the interlayer shear modes are much more sensitive to the interlayer coupling strength, hence they are only prominent at twist angles very close to 0 and 60°.[24, 25] As shown in Figure 2b, the calculated breathing mode frequencies of the S/S heterobilayer display the same pattern as the measurements. They exhibit a local maximum frequency near 30° that decreases by around 1 cm$^{-1}$ as the twist angle decreases to 13.2° or increases to 46.8°. A similar trend is observed for the Se/S interface. Additionally, the calculations in Figure 2b reproduce the experimental red-shift of the breathing mode of the Se/S heterobilayer as compared to the S/S heterobilayer (Figure 1d).

The characteristics of the low-frequency Raman modes can be explained by the interlayer distance $d_{Mo-Mo}$, defined as the average vertical distance between the Mo atoms in the MoSSe and MoS$_2$ layers. The lower frequency of breathing modes in heterostructures with $\theta = 13.2$ and 46.8° is attributed to the larger interlayer distance as compared to the other twist angles, for both S/S and Se/S heterobilayers (Figure 2c). The influence of the interlayer distance $d_{Mo-Mo}$ on the interlayer interaction is most obvious when comparing untwisted and twisted structures. Specifically, the interlayer distance $d_{Mo-Mo}$ for $\theta = 5-55°$ increases by around 10% compared to $d_{Mo-Mo}$ for $\theta = 0$ and 60° (Figure 2c), leading to a decrease in breathing mode frequencies by around 8 cm$^{-1}$ (Figure 2b).

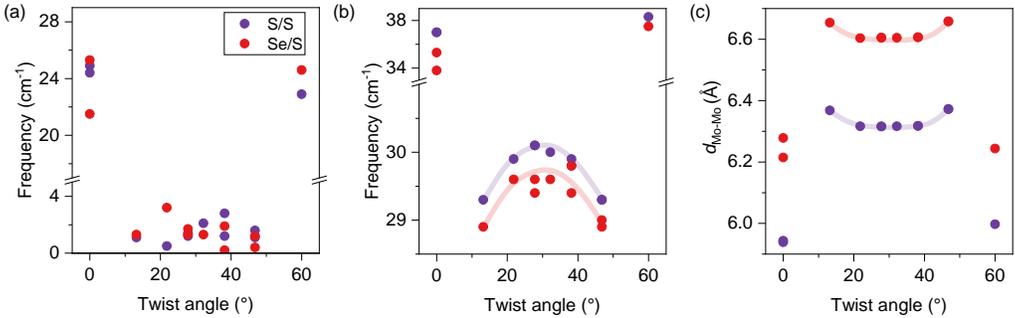

**Figure 2**. The DFT calculated low-frequency Raman modes and interlayer distance of MoSSe/MoS$_2$ with S/S and Se/S interfaces. The calculated (a) interlayer shear mode and (b) interlayer breathing mode frequencies for different twist angles. (c) The calculated interlayer distance for different twist angles. The curves in (b-c) are guides to the eye.

**Interlayer Coupling Revealed by High-Frequency Raman Modes**

Interfacial configurations not only alter the low-frequency Raman modes, but also the high-frequency Raman modes. The MoSSe/MoS$_2$ with the S/S interface displays the E and A$_1$ modes of MoSSe at 354 and 290 cm$^{-1}$, and the E' and A$_1$' modes of MoS$_2$ at 383 and 404 cm$^{-1}$ (Figure 3a and Figure S4 for different twist angles). The irreducible representations of the lattice vibrations are determined by the point group symmetry of MoSSe (C$_{3v}$) and MoS$_2$ (D$_{3h}$) monolayers. In Figures 3b and 3c, we summarize the average frequencies of the S/S and Se/S heterobilayers with various twist angles and include the same information for 2H and 3R MoSSe/MoS$_2$ heterobilayers fabricated by direct synthesis.[8] The heterostructures also possess two bulk vibrational modes that are sensitive to the number of layers. These modes, namely the A$_{1g}^2$ mode at 441 cm$^{-1}$ for MoSSe and at 470 cm$^{-1}$ for MoS$_2$, are active only in the heterostructure but not in either MoSSe or MoS$_2$ monolayers (Figure S5). Since the A$_{1g}^2$ modes are only Raman-active in multilayer TMDs, we use the notation for the D$_{3d}$ point group of 2H MoS$_2$ bilayers.[29]

Although the high-frequency Raman modes are not as susceptible to the twist angles as low-frequency Raman modes, their frequencies are modified by the stacking symmetry and the species of atoms at the interface. We first analyze the effect of stacking symmetry by comparing directly synthesized and transfer stacked samples. The directly synthesized 2H and 3R samples are bilayer MoS$_2$ that have gone through selenization and have intrinsically ideal S/S interfaces. In comparison with the 2H and 3R heterobilayers, the MoSSe E and A$_1$ modes of twisted heterostructures (both S/S and Se/S heterobilayers) red-shift by 1-4 cm$^{-1}$, as shown in Figure 3b. This can be attributed to the relaxation of in-plane compressive strain in the MoSSe layer after transfer stacking.[8] Due to the slight lattice mismatch between MoSSe and MoS$_2$ monolayers, the as-grown MoSSe monolayer is confined by both the original MoS$_2$ lattice and the substrate. This compressive strain is partially relaxed when the MoSSe is detached from the substrate and transferred onto another as-grown MoS$_2$. In contrast, for the high-frequency modes of MoS$_2$ in Figure 3c, the E' mode blue-shifts while the A$_1$' mode red-shifts for the twisted MoSSe/MoS$_2$ compared to the 2H and 3R heterobilayers. This is not explained by strain relaxation but rather by the dependence on layer thickness. For few-layer TMD, the E' and A$_1$' modes shift to higher and lower frequencies, respectively, when the number of layers decreases. This again indicates that the twisted MoSSe/MoS$_2$ possesses weaker interlayer coupling than MoSSe/MoS$_2$ with 2H and 3R stackings.

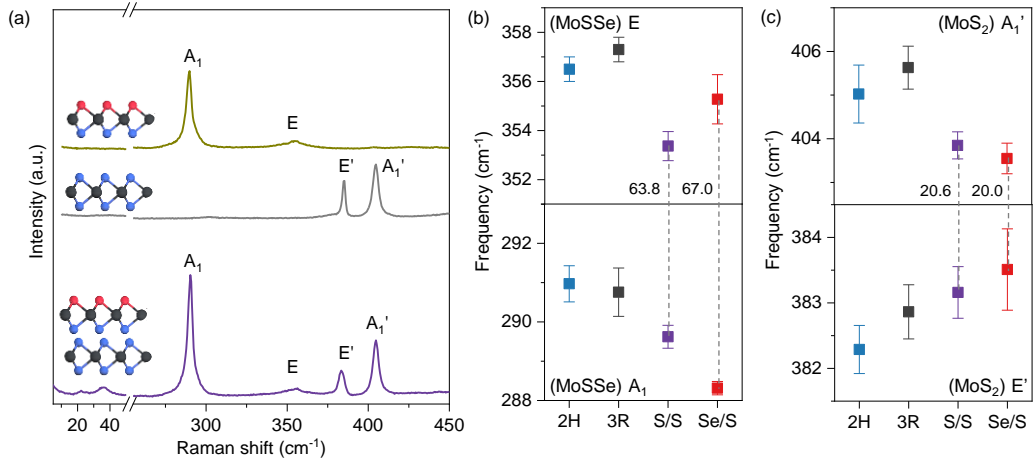

**Figure 3**. High-frequency Raman modes of MoSSe/MoS$_2$. (a) Raman spectra of monolayer MoSSe, monolayer MoS$_2$, and MoSSe/MoS$_2$ with S/S interface. The frequencies of (b) the E and A$_1$ modes of MoSSe and (c) the E' and A$_1$' modes of MoS$_2$ in 2H and 3R stackings by direct synthesis (S/S interface),[8] and in twisted heterostructures with S/S and Se/S interfaces produced by transfer stacking. The squares and the error bars are the mean value and the standard deviation, respectively. The frequency separations for twisted S/S and Se/S heterostructures are labeled in (b-c).

Meanwhile, the influence of interface composition (S/S and Se/S) on the interlayer coupling is reflected by the frequency separation of high-frequency Raman modes. For few-layer MoS$_2$, the frequency separation of the E' and A$_1$' modes $\Delta\omega_M = \omega_{A_1'} - \omega_{E'}$ is an indicator of the layer thickness.[30] The typical $\Delta\omega_M$ for MoS$_2$ monolayer (bilayer) is around 19 (22) cm$^{-1}$.[30, 31] Therefore, a value of $\Delta\omega_M$ closer to 19 cm$^{-1}$ indicates a weaker interlayer coupling in the heterostructure so that the MoS$_2$ behaves more like a separated monolayer. As expected, the Se/S heterostructure has a smaller $\Delta\omega_M$ than the S/S heterostructure, namely 20.0 cm$^{-1}$ compared to 20.6 cm$^{-1}$. This suggests that the Se/S interface has weaker interlayer coupling compared to the S/S interface, which agrees with the analysis of the low-frequency Raman modes. The correlation between frequency separation and interlayer coupling is more evident by comparing the 2H and 3R heterobilayers to the twisted samples. The stronger interlayer interaction in 2H and 3R heterostructures yields a $\Delta\omega_M$ of 22.7 cm$^{-1}$, which is by around 2 cm$^{-1}$ higher than the $\Delta\omega_M$ of twisted heterostructures. It is worth noting that the E' and A$_1$' mode frequencies do not show a clear dependence on the twist angles, because they are influenced by the strain and defect conditions of the composite monolayers.[32, 33] However, $\Delta\omega_M$ for the twisted S/S heterostructure is higher for twist angles close to 0/60° than the other angles by around 0.6 cm$^{-1}$ (Figure S6). This twist-angle dependence is similar to observations in twisted bilayer MoS$_2$[24, 34] as well as twisted WS$_2$/MoS$_2$ heterostructures.[35]

Similarly, the frequency separation of the E and A$_1$ modes of Janus MoSSe, $\Delta\omega_J = \omega_E - \omega_{A_1}$ conveys the strength of interlayer coupling. According to DFT calculations of 2H MoSSe with Se/S interfaces (Figure S7), $\Delta\omega_J$ is 64.6 and 60.9 cm$^{-1}$ for MoSSe monolayers and bilayers, respectively, demonstrating that the interlayer interaction results in a smaller value of $\Delta\omega_J$. As shown in Figure 3b, $\Delta\omega_J$ of the S/S heterobilayer (63.8 cm$^{-1}$) is smaller than that of the Se/S heterobilayer (67.0 cm$^{-1}$) by around 3 cm$^{-1}$. This reduced $\Delta\omega_J$ suggests a stronger interfacial interaction in S/S heterostructure, which is consistent with the conclusion from the analysis of $\Delta\omega_M$. DFT calculations reproduce the frequency separations as shown in Figure S8. The DFT calculated $\Delta\omega_J$ for S/S heterostructure is smaller by 2.4 cm$^{-1}$ than that for Se/S heterobilayer. Both our experiments and DFT calculations show that the E (MoSSe) and E' (MoS$_2$) modes have larger frequency variations among twist angles compared to the A$_1$ (MoSSe) and A$_1$' (MoS$_2$) modes (Figure 3b-c and Figure S8). It suggests that the E and E' modes are more affected by in-plane atomic arrangement since they represent in-plane vibrations of the chalcogen atoms, and the in-plane atomic arrangement indeed undergoes structural renormalization when forming twisted heterostructures.

**PL Quenching Due to Interfacial Charge Transfer**

In layered transition metal dichalcogenides, interlayer and intralayer excitonic behaviors are modulated by a wide range of physical parameters including interlayer coupling, charge density distribution, band structure, and defects. The intralayer excitons of the as-grown monolayer MoS$_2$ are at 1.85 and 2.00 eV, while the transferred MoSSe has the exciton energy of 1.74 eV (Figure 4a). In the twisted heterostructures, the S/S and Se/S heterobilayers have two PL peaks at the same energies around 1.84 and 1.97 eV (Figure 4b) corresponding to the A and B excitons of MoS$_2$, while the intralayer exciton of MoSSe is not observable due to intrinsically low PL intensity. The slight decreases of the A and B exciton energies of MoS$_2$ are consistent with the shifts of the DFT imaginary part of permittivity of MoS$_2$ under tensile strain (Figure S9). Unlike low-frequency Raman modes, the experimentally measured PL energy of intralayer excitons does not explicitly depend on the twist angle (Figure S10). The PL intensity of A and B excitons of MoS$_2$ in the heterostructure is reduced by around 5 and 2 times, respectively, compared to the A and B excitons in the MoS$_2$ monolayer (Figure 4a), which could be explained by the PL quenching due to interfacial charge transfer.[36, 37] To quantify the PL quenching effect, we define the PL intensity ratio of the A and B excitons as the ratio of the PL intensity in the MoSSe/MoS$_2$ heterobilayer to the PL intensity in the constituting MoS$_2$ monolayer. A smaller PL intensity ratio corresponds to a stronger PL quenching effect. As shown in Figure 4c, the PL intensity ratio of the A exciton is around 0.1 and 0.2 for S/S and Se/S heterobilayers, respectively. The smaller PL intensity ratio of the A exciton in the S/S heterostructure suggests greater charge transfer at the interface. Upon photo-excitation, the excited electrons flow from the conduction band of MoSSe to MoS$_2$, while holes transition from the valence band of MoS$_2$ to MoSSe. Thus the recombination through intralayer excitons is impaired by the electron-hole separation across layers. On the other hand, the PL intensity ratio of the B exciton is around 0.5 for both S/S and Se/S heterobilayers, which is significantly larger than the PL intensity ratio of the A exciton.

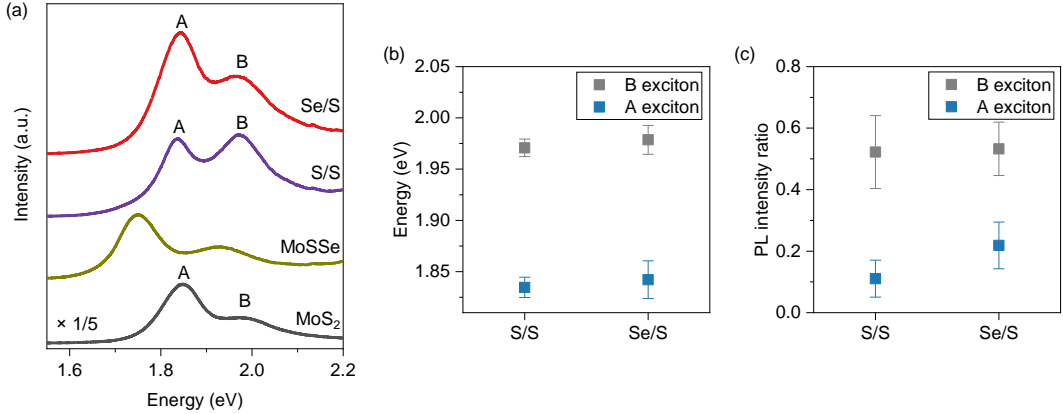

**Figure 4.** PL quenching of MoSSe/MoS$_2$. (a) PL spectra of MoS$_2$ monolayer, MoSSe monolayer, and twisted heterobilayers with S/S and Se/S interfaces. The PL intensity of the MoS$_2$ monolayer is scaled down by a factor of 5. (b) The PL energy of the A and B excitons of twisted heterobilayers with S/S and Se/S interfaces. (c) The PL intensity ratio defined as the ratio between the PL intensities in the heterobilayers and the corresponding PL intensities in the MoS$_2$ monolayer, for both A and B excitons and both S/S and Se/S heterobilayers.

The photogenerated charge transfer can be affected by the intrinsic electric field at the interface. Since the work function difference of the constituent layers affects the charge redistribution, we calculated the work functions of S/S and Se/S heterostructures as shown in Figures 5a-b. The electrostatic potential difference between the two chalcogen layers is 0.77 eV in MoSSe. It imposes different work functions when different chalcogens are in contact with MoS$_2$ causing an electrostatic potential difference of 0.76 and −0.65 eV across S/S and Se/S heterostructures, respectively (Figures 5a-b). This electrostatic difference leads to a built-in interfacial electric field with opposite directions in S/S and Se/S heterostructures. For S/S (Se/S) stacking, the intrinsic interfacial electric field points from MoS$_2$ to MoSSe (from MoSSe to MoS$_2$) as shown by the DFT charge density difference in Figures 5c-d. For Se/S, there is more electron accumulation close to the sulfur atoms of the MoS$_2$ layer and more electron depletion close to the selenium atoms of the MoSSe layer. Electrons also tend to accumulate more near the Mo atoms in MoS$_2$ as compared to the Mo atoms in MoSSe for Se/S stacking (Figure 5d). In contrast, for S/S stacking the Mo atoms in MoS$_2$ have less electron accumulation than the Mo atoms in MoSSe (Figure 5c inset). We also conjecture that the built-in interfacial electric field has a larger amplitude in Se/S than S/S in addition to the opposite direction. This is because the charge density distribution near the sulfur atoms of MoS$_2$ at the interface is more uneven in Se/S stacking than S/S stacking. For S/S heterobilayers, the intrinsic electric field pointing from MoS$_2$ to MoSSe promotes electron transfer from MoSSe to MoS$_2$ along with hole transfer from MoS$_2$ to MoSSe. On the other hand, electron transfer to MoS$_2$ at the Se/S interface is opposed by the intrinsic electric field. Therefore, the PL quenching of intralayer excitons could be more effective in the S/S heterostructure than in the Se/S heterostructure because of the charge transfer facilitated by the built-in electric field. The contrasting PL intensity ratio of A and B excitons can also be easily understood by the fact that the A exciton resides closer to the band extrema than the B exciton and is more influenced by interfacial charge transfer.

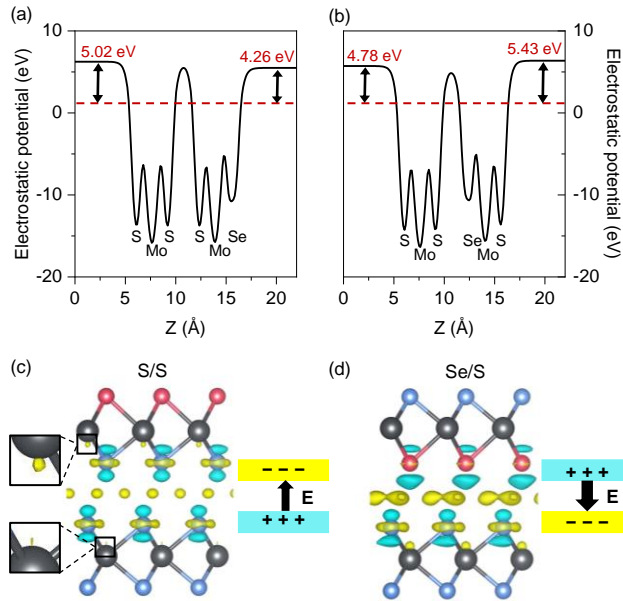

**Figure 5**. Charge density distribution of Janus MoSSe/MoS$_2$ heterostructures. (a-b) Local electrostatic potential (ionic and Hartree contributions) for (a) S/S and (b) Se/S heterostructures. The red dashed line represents the Fermi level. (c-d) Charge density differences for (c) S/S and (d) Se/S heterostructures. The cyan and yellow areas represent electron depletion and accumulation compared to the isolated monolayers. The black arrows label the direction of the electric field at the interface.

In addition to the built-in interfacial electric field, interlayer coupling and band hybridization can contribute to interfacial charge transfer. As we have demonstrated by the phonon modes, the S/S interface exhibits stronger interlayer coupling than the Se/S interface. Our DFT band diagrams in Figure 6 also support this conclusion. By comparing the band structures of the heterobilayers (Figures 6a-b) with the superimposed band structures of individual monolayers (Figures 6c-d), we observe that the MoS$_2$ layer in S/S heterostructure becomes an indirect bandgap semiconductor similar to a conventional TMD bilayer, while the MoS$_2$ layer in Se/S heterostructure retains its direct bandgap. This indicates that heterostacking and the resulting interlayer coupling significantly affect the band structure of the S/S heterobilayer. In contrast, Se/S stacking does not noticeably alter the band structures of individual layers. Thus, the PL quenching of MoS$_2$ can be partially explained by the direct-to-indirect bandgap transition. Yuan *et al.* studied the PL behavior of MoS$_2$/WS$_2$ heterostacks and observed a stronger PL quenching for WS$_2$ than MoS$_2$.[37] They explained the difference in PL quenching using DFT calculations showing that the direct bandgap is maintained in MoS$_2$, while the WS$_2$ layer in the heterobilayer has a transition from a direct bandgap to an indirect bandgap. This is consistent with our observation of a stronger PL quenching of the MoS$_2$ A exciton for S/S stacking as compared to Se/S stacking. In addition, the S/S heterostructure exhibits more band hybridization between the conduction bands close to the band extrema and between the valence bands at the $\Gamma$ point, compared to the corresponding bands in the Se/S heterostructure (Figures 6a-b). It has been demonstrated for PTCDA molecule/WSe$_2$ heterojunctions that a higher degree of hybridization of the WSe$_2$ conduction band with PTCDA unoccupied states contributes to a larger amount of charge transfer.[38] Therefore, the greater charge transfer in S/S heterostructures can also be understood by the band hybridization resulting from the interlayer coupling. One would intuitively think that the interlayer distance has a substantial influence on the charge transfer dynamics. Nonetheless, a time-dependent DFT study on MoS$_2$/WS$_2$ indicates that the charge transfer process is dominated by the coupling between the electronic states rather than interlayer distance.[39] Thus, we conjecture that the interlayer separation of MoSSe/MoS$_2$ with S/S and Se/S interfaces is not the key factor for interfacial charge transfer. This is consistent with our observation that the PL intensity ratio does not show a clear dependence on the twist angle for either S/S or Se/S heterobilayers (Figure S11).

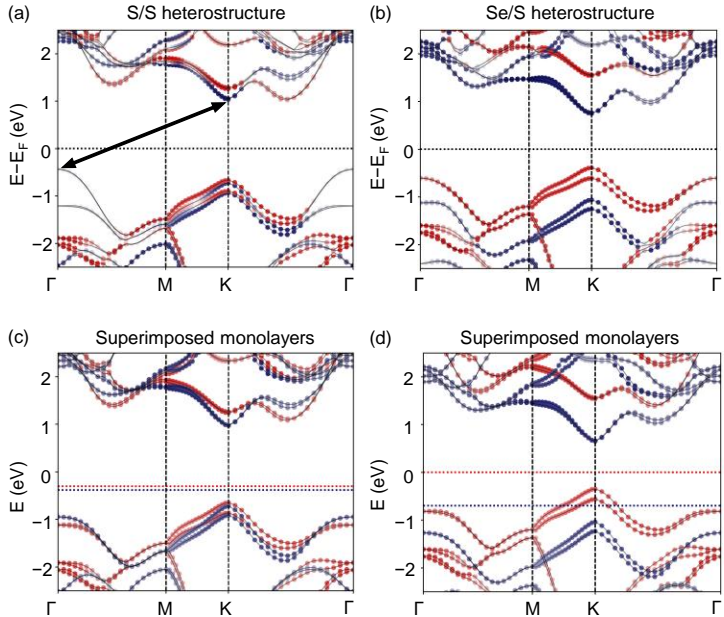

**Figure 6**. Electronic band diagrams of (a) S/S and (b) Se/S bilayer heterostructures, and (c-d) superimposed band structures of individual monolayers. Band energies were calculated with a hybrid functional including spin-orbit coupling. The red and blue dots indicate the projection onto MoSSe or MoS$_2$ layers, respectively, with fainter colors indicating a higher degree of band hybridization. The black arrow in (a) indicates possible indirect transition for S/S heterostructure. The band energies of the separate MoSSe and MoS$_2$ layers in (c-d) are rigidly shifted in energy so that the valence and conduction band extrema match with those of the heterostructures. The Fermi level for each layer is indicated by a horizontal dotted line of the same color.

Finally, we comment briefly on the future exploration of Janus TMDs. Our conclusions about the charge transfer properties are inferred from the PL quenching behaviors. To directly probe the charge transfer pathways, pump-probe transient absorption can provide details on the excited states in specific constituents with temporal resolutions. In addition to the intralayer excitons, it is important to investigate the interlayer excitonic transitions, including the exciton lifetime and the relaxation channels, that might yield further insights into the effects of the intrinsic dipole moment. A computational study on the MoSSe/WSeTe by Zhou *et al.* argued that charge transfer is favored at the Te/Se interfaces while prohibited at the Te/S interface, which is confirmed by their non-adiabatic molecular dynamics calculations of the electron transfer lifetimes.[40] From an experimental viewpoint, the recent developments in the synthesis of Janus TMD heterostructures[9,10] will enable further investigation of Janus multilayers and the unique properties of their van der Waals interfaces.

**CONCLUSIONS**

In summary, we investigated the capability of Janus MoSSe to tune the van der Waals coupling with monolayer $MoS_2$ and the subsequent phononic and excitonic properties. We studied the detailed correlation of those properties with the in-plane twist angle and the out-of-plane intrinsic dipole moment of Janus MoSSe. Similar to conventional TMD bilayers, the Janus heterobilayers exhibit interlayer breathing mode frequencies that peak at $\theta = 30°$ for samples with intermediate twist angles between 5 and 55°. We also demonstrate that the asymmetry of Janus MoSSe has fundamental implications for the interlayer interactions. In particular, the S/S heterobilayer possesses a smaller interlayer separation and thus stronger interfacial coupling than the Se/S interface. Such a difference is not only revealed by the interlayer breathing mode frequency but also reflected by the frequency separation of the high-frequency optical phonons. Moreover, the PL quenching of $MoS_2$ intralayer excitons in $MoSSe/MoS_2$ heterostructure is influenced by the interfacial charge transfer, which is different for S/S and Se/S interfaces. The stronger quenching in the S/S heterobilayer is attributed to the enhanced charge transfer caused by the interfacial electric field resulting from charge density redistribution and band hybridization. Our spectroscopic study, integrated with DFT calculations, sheds light on the dipole-modulated charge transfer and the asymmetry-induced phonon properties that are only accessible in Janus TMD-related heterostructures. The fundamental understanding of the microscopic mechanical, electronic, and optical behaviors of Janus heterostructures provides a facile yet effective approach to manipulating van der Waals interactions and the corresponding electronic and optical properties.

**METHODS**

***Sample fabrication***: $MoS_2$ monolayer was synthesized using a seeding promoter perylene-3,4,9,10-tetracarboxylic acid tetrapotassium salt (PTAS) assisted CVD method on $SiO_2/Si$ substrate. Then, the $MoS_2$ on $SiO_2/Si$ was cut into two pieces, one of which was selenized into MoSSe using the method reported in Ref. 8 and 9 and stacked onto the other half of $MoS_2$ to form Janus $MoSSe/MoS_2$ heterostructures. For the S/S heterobilayer, the MoSSe was wet-transferred onto the as-grown $MoS_2$. For the Se/S heterobilayer, the MoSSe was first flipped over and then transferred onto the $MoS_2$. Both kinds of heterostructures were annealed at 200 ºC in vacuum for 2 hours after removing the PMMA layer with acetone.

***Optical spectroscopy***: The Raman and PL spectroscopies were performed on a Horiba LabRAM HR Evolution spectrometer with an 1800/mm grating in backscattering geometry. The 532 nm laser was employed as the excitation wavelength and focused on the sample with a 100× objective. An ultra-low frequency module was incorporated to obtain a spectral limit down to 10 cm$^{-1}$. The spectral resolution by 1800/mm grating is around 0.2 cm$^{-1}$. The Raman frequencies were obtained from the least square fitting based on the Levenberg–Marquardt algorithm using Lorentz functions.

*Density functional theory*: The heterobilayer structures were modeled using commensurate supercells in a slab geometry, with a vacuum space of 16 Å. All DFT calculations were performed with the VASP code[41, 42] using a plane-wave energy cutoff of 500 eV with PAW pseudopotentials.[43, 44] Calculation for twisted cells used the local density approximation (LDA) exchange-correlation functional. The atomic structures were relaxed until the forces on each atom were below 10$^{-4}$ eV/Å. The Γ-point phonon frequencies were calculated using the density functional perturbation theory approach. The Brillouin zone (BZ) was sampled by a 5×5×1 Γ-centered grid for both relaxation and phonon calculations, except for the largest cells, 13.2° and 46.8°, where the k-point sampling was reduced to 4×4×1 for the phonon calculation. The electronic band structures included spin-orbit coupling and were calculated using the HSE hybrid functional with an energy cutoff of 280 eV and a screening parameter of μ = 0.4 Å$^{-1}$ that was optimized by Zahid *et al.*[45] to match the experimental band gap in MoS$_2$ for single layers, bilayers, and bulk. The electrostatic potential was calculated with a dipole correction and includes ionic and Hartree terms but no exchange-correlation contributions. The energy-dependent dielectric tensor including excitonic effects (Figure S9) was calculated by solving the Bethe-Salpeter equation after a GW0 calculation of the primitive monolayer cell with 3 atoms. Spin-orbit interactions were included, the energy cutoff was 280 eV, and the BZ was sampled with an 18×18×1 k-point grid. Four occupied and eight valence bands were included. Calculations in Figure S2 were performed assuming a lattice constant of 3.1 Å and a vacuum space of 22 Å. The BZ was sampled by a 15×15×1 Γ-centered grid. For a given stacking configuration, the in-plane atomic positions were kept fixed, while the out-of-plane positions were allowed to relax until the forces on each atom were below 5×10$^{-4}$ eV/Å.

## ASSOCIATED CONTENT

**Supporting Information**

The Supporting Information is available free of charge at https://pubs.acs.org/doi/

Experimental Stokes and anti-Stokes low-frequency Raman spectra (Figure S1); DFT calculated relative energy of high-symmetry stacking heterostructures (Figure S2); atomic configurations considered in DFT calculations for twist S/S heterostructures (Figure S3); experimental high-frequency Raman spectra (Figure S4-5) and frequency separations (Figure S6); DFT calculated high-frequency Raman modes for MoSSe multilayers (Figure S7) and twist heterostructures (Figure S8); DFT calculated imaginary part of the dielectric constant of MoS$_2$ monolayers (Figure S9); experimental PL energy and intensity ratio for different twist angles (Figure 10-11).

## ACKNOWLEDGEMENT

K.Z. and S.H. acknowledge the support from National Science Foundation Award No. DMR-2011839 through the Penn State MRSEC–Center for Nanoscale Science. Y.G. and J.K. acknowledge the support from the U.S. Department of Energy, Office of Science, Basic Energy Sciences, award number DE-SC0020042 for the synthesis and fabrication of the Janus TMD and heterostructures. D.T.L. acknowledges helpful DFT discussions with Michele Pizzochero and Oscar Grånäs. D.T.L. and Z.Z. are supported by NSF DMREF Grant No. 1922165 and the NSF STC Center for Integrated Quantum Materials, NSF Grant No. DMR-1231319. D.T.L. is also supported by DOE Basic Energy Science Award No. DE-SC0019300. Z.Z. is also supported by ARO MURI Grant No. W911NF-14-0247. The DFT calculations used resources of the National Energy Research Scientific Computing Center (NERSC), a U.S. Department of Energy Office of Science User Facility located at Lawrence Berkeley National Laboratory, operated under Contract No. DE-AC02-05CH11231 and the FASRC Cannon cluster supported by the FAS Division of Science Research Computing Group at Harvard University. S.F. is supported by a Rutgers Center for Material Theory Distinguished Postdoctoral Fellowship. S.H. acknowledges the support from the National Science Foundation under Grant No. ECCS-1943895.

**For Table of Contents Only**

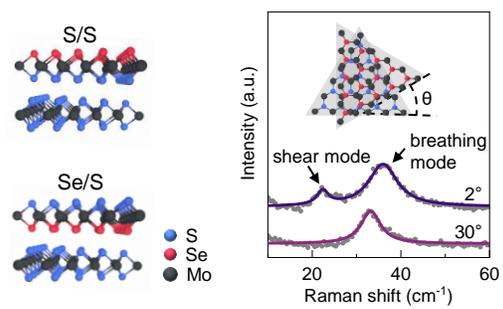

# Supporting Information

Spectroscopic Signatures of Interlayer Coupling in Janus MoSSe/MoS$_2$ Heterostructures

Kunyan Zhang[1], Yunfan Guo[2,*], Daniel T. Larson[3], Ziyan Zhu[3], Shiang Fang[4], Efthimios Kaxiras[3,5], Jing Kong[2,*], Shengxi Huang[1,*]

[1]Department of Electrical Engineering, The Pennsylvania State University, University Park, Pennsylvania 16802, United States

[2]Department of Electrical Engineering and Computer Science, Massachusetts Institute of Technology, Cambridge, Massachusetts 02139, United States

[3]Department of Physics, Harvard University, Cambridge, Massachusetts 02138, United States

[4]Department of Physics and Astronomy, Center for Materials Theory, Rutgers University, Piscataway, New Jersey 08854, United States

[5]John A. Paulson School of Engineering and Applied Sciences, Harvard University, Cambridge, Massachusetts 02138, United States

*Corresponding author

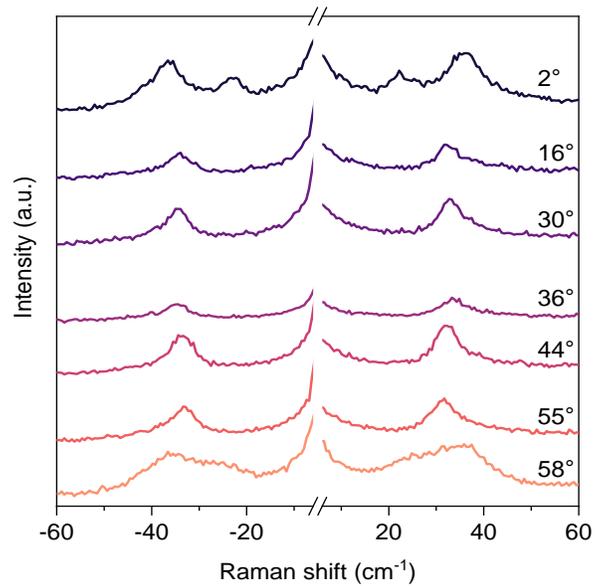

**Figure S1**. Stokes and anti-Stokes Raman spectra of the low-frequency modes with different twist angles for MoSSe/MoS$_2$ with S/S interfaces. The spectra are shifted in the *y*-axis.

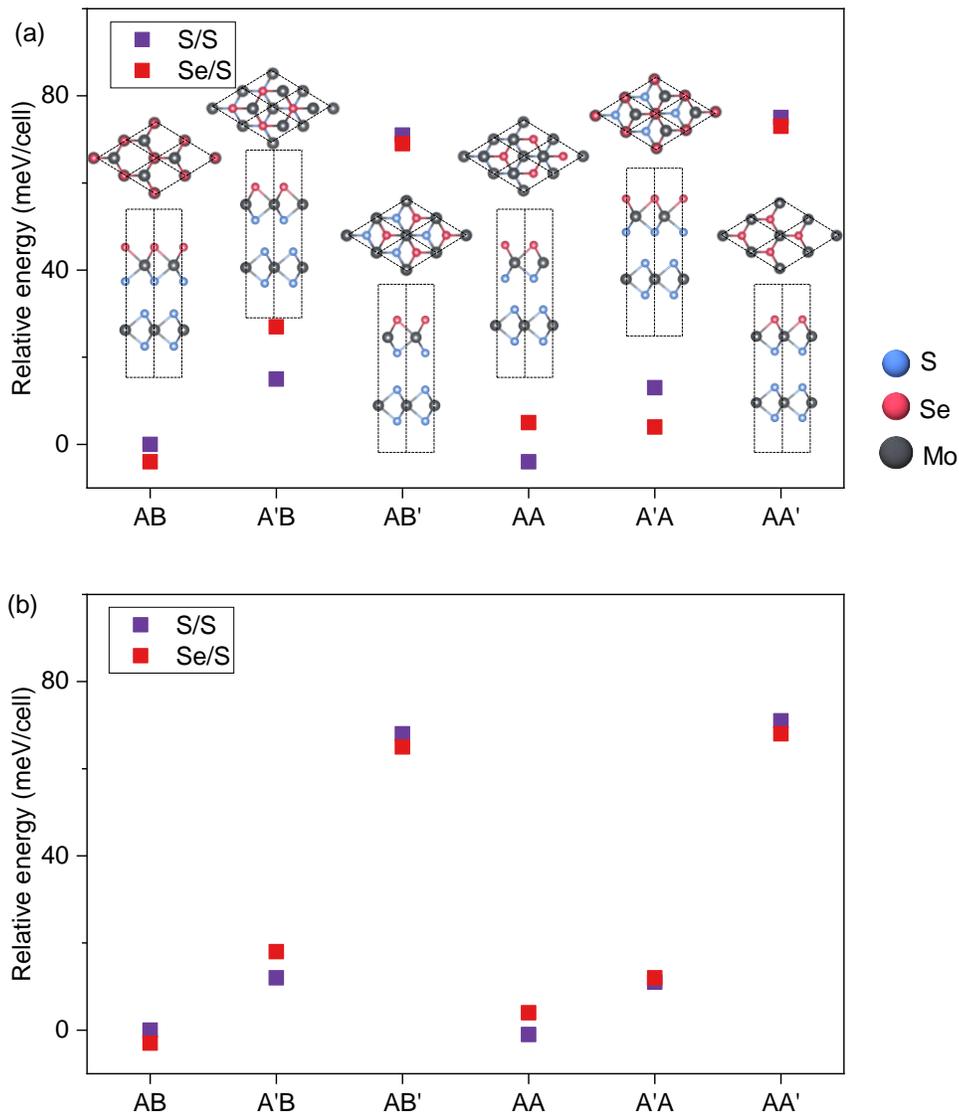

**Figure S2**. The relative energy of MoSSe/MoS$_2$ heterostructures with high-symmetry stackings by local density approximation (LDA) exchange-correlation functional (a) without and (b) with van der Waals correction. The values are relative to the S/S heterostructure with AB stacking. Insets in (a) are the top view and side view of the respective stacking geometry for S/S heterobilayers. The AB and AA stackings correspond to the 2H and 3R stackings, respectively. The A'A is an additional high-symmetry stacking to the five high-symmetry stackings of MoS$_2$ bilayers.

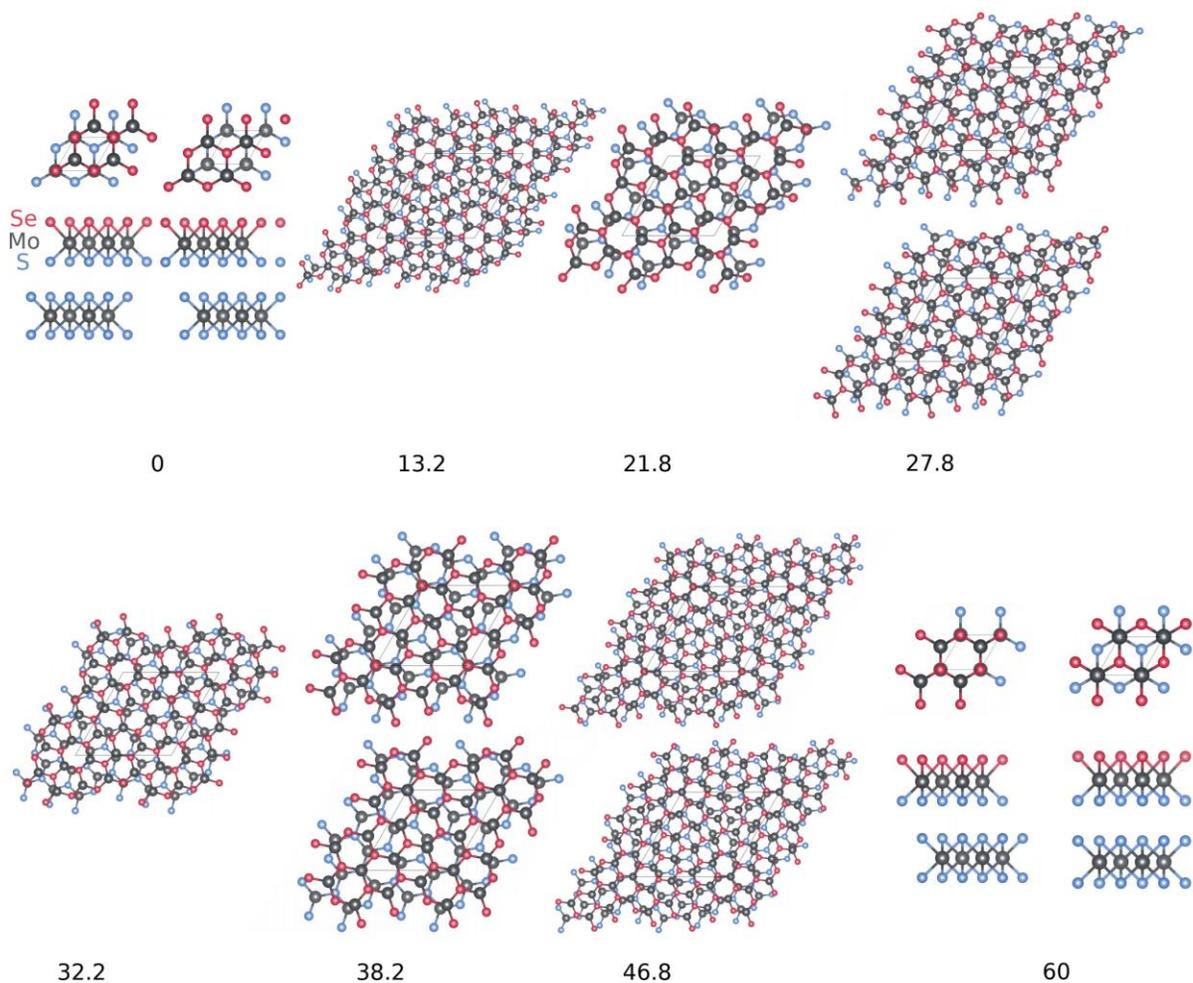

**Figure S3**. Atomic configurations for density functional theory (DFT) calculation of the Raman modes for S/S heterobilayers. The twist angles considered include 0°, 13.2°, 21.8°, 27.8°, 32.2°, 38.2°, 46.8°, 60°, in which 0°, 27.8°, 38.2°, 46.8°, 60° encompass two stacking configurations that are related by in-plane translations. The stacking geometries considered for Se/S heterobilayers are similar to the above figure, but with the sulfur and selenium atoms in the Janus MoSSe layer swapped.

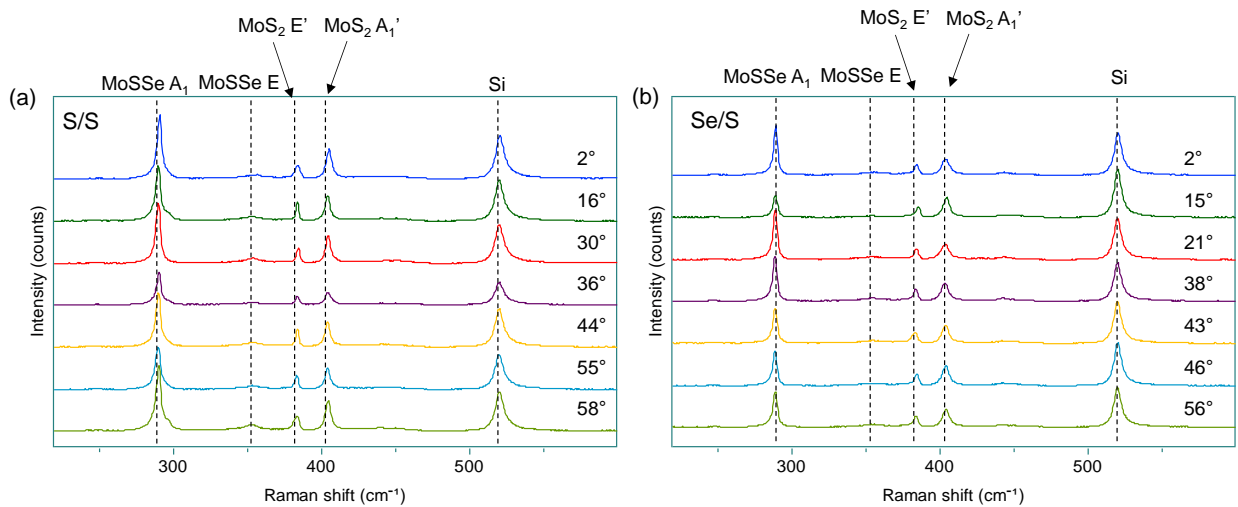

**Figure S4**. High-frequency Raman spectra of S/S and Se/S heterostructures at representative twist angles. The spectra are shifted in the *y*-axis direction.

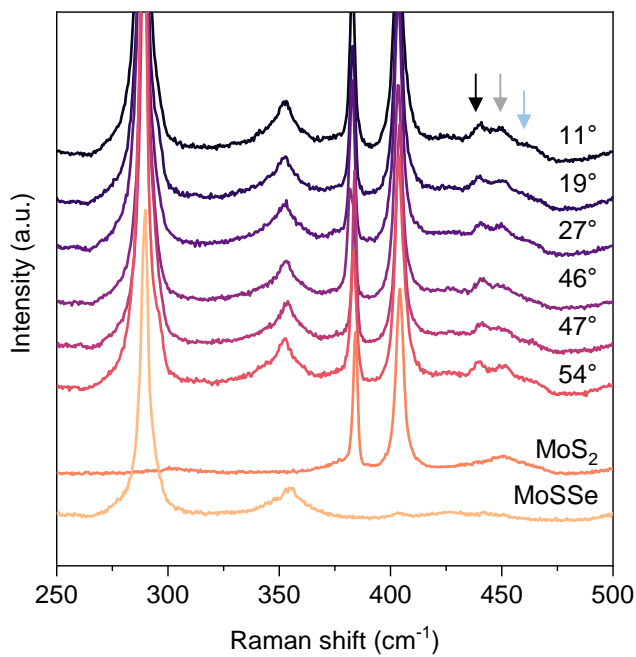

**Figure S5**. The experimental high-frequency Raman spectra of S/S heterobilayer. The three arrows around 450 cm$^{-1}$ represent the bulk vibrational $A_{1g}^2$ mode of Janus MoSSe (black arrow), the second-order 2LA(M) mode of $MoS_2$ (grey arrow), and the bulk vibrational $A_{1g}^2$ mode of $MoS_2$ (blue arrow). The Raman spectra of $MoS_2$ and MoSSe monolayers are included for comparison. The spectra are shifted in the *y*-axis.

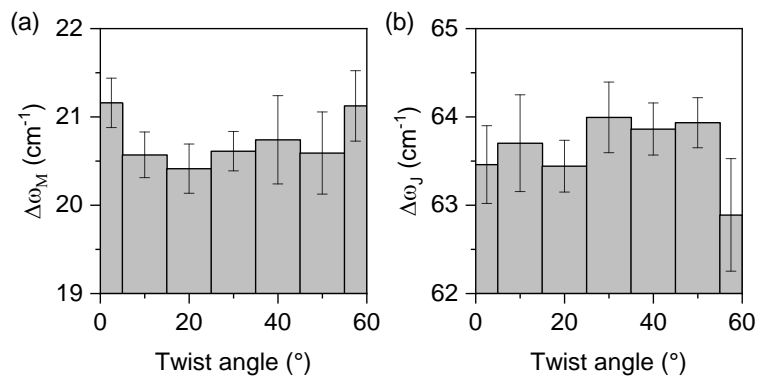

**Figure S6**. The experimental frequency separations of the S/S heterobilayer. (a) The separation of the $A_1'$ and $E'$ modes of $MoS_2$ and (b) the E and $A_1$ modes of MoSSe. The bins are the mean values, and the error bars are the standard deviations for twist angles of 0-5°, 5-15°, 15-25°, 25-35°, 35-45°, 45-55°, 55-60°.

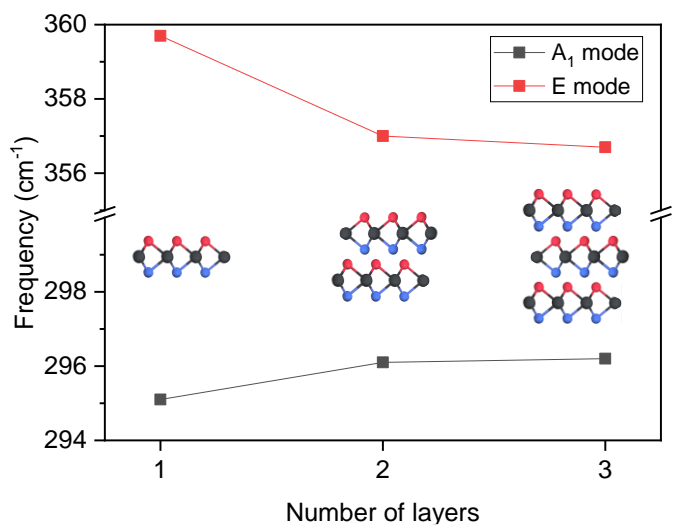

**Figure S7**. DFT calculated high-frequency Raman modes of few-layer MoSSe. The bilayer and trilayer have 2H stacking and the interfaces are composed of the selenium from one layer and the sulfur from the other layer.

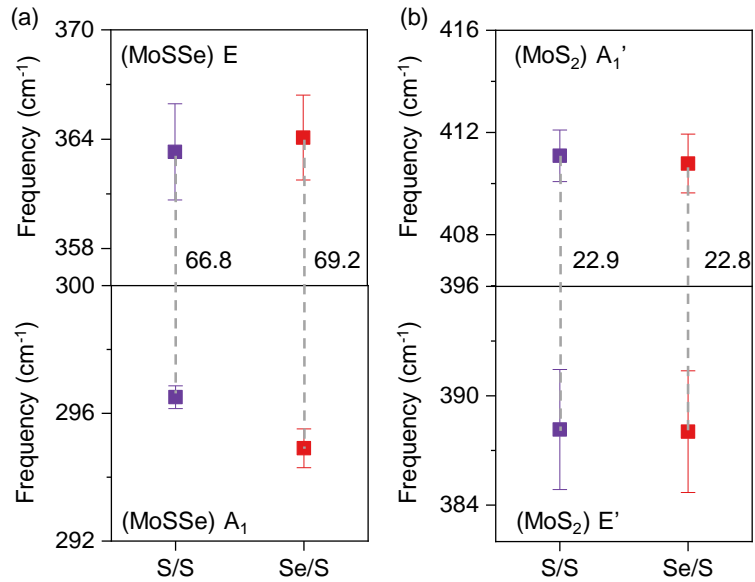

**Figure S8**. DFT calculated high-frequency Raman modes. (a) The E and $A_1$ modes of MoSSe and (b) the $A_1$' and E' modes of $MoS_2$ for MoSSe/$MoS_2$ heterostructures with S/S and Se/S interfaces. The squares are the mean values for different twist angles. The error bars are the standard deviations. The frequency separations are labeled in the figure.

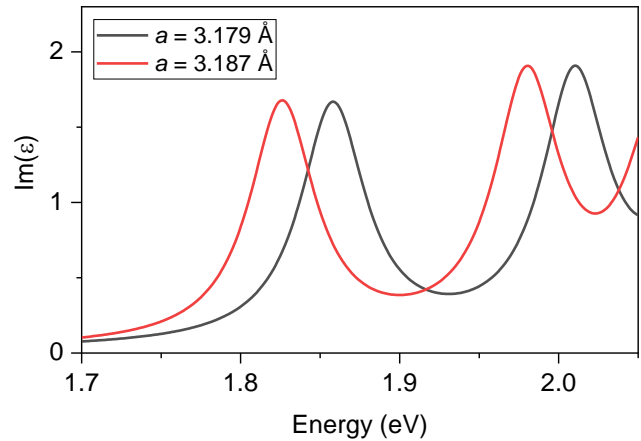

**Figure S9**. GW0+BSE calculation of the imaginary part of the dielectric constant of the $MoS_2$ monolayer with a lattice constant $a$ = 3.179 and 3.187 Å, broadened by a Lorentzian with a width of 0.05 eV and rescaled vertically.

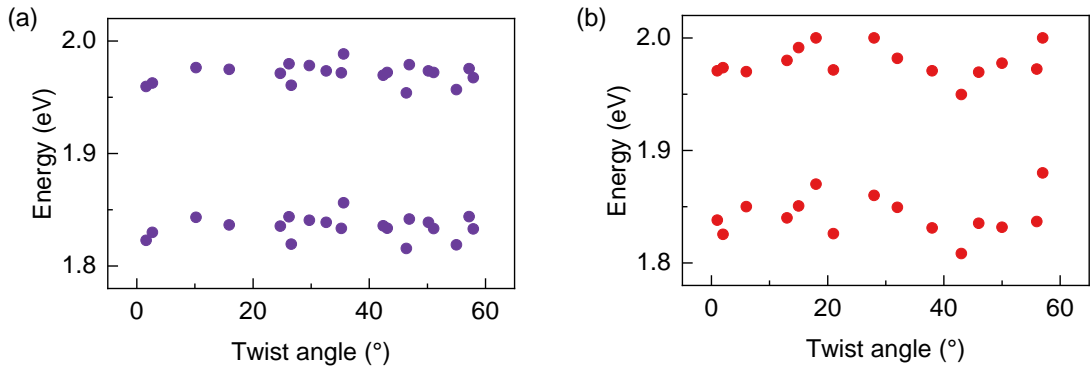

**Figure S10**. The experimental photoluminescence (PL) energy of the $MoS_2$ A and B excitons of (a) S/S and (b) Se/S heterobilayers.

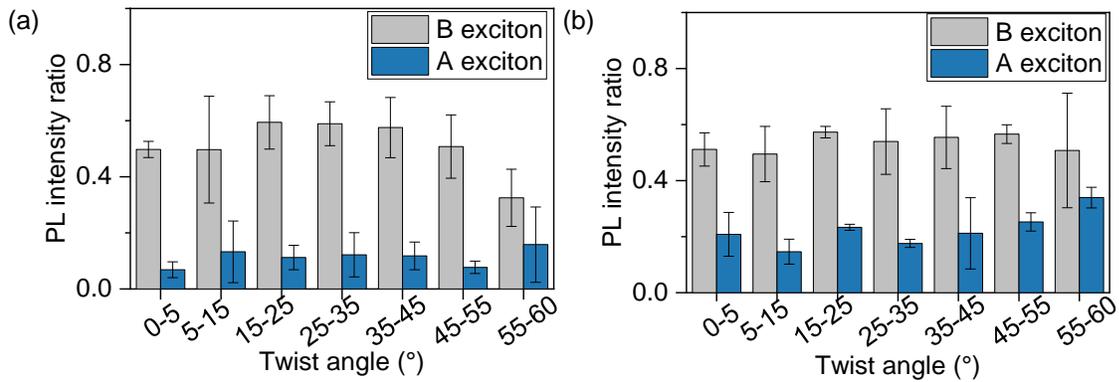

**Figure S11**. The experimental PL intensity ratio of the excitonic peak intensities between the heterobilayer and the corresponding $MoS_2$ monolayer for both A and B excitons and for (a) S/S and (b) Se/S heterobilayers as a function of twist angles. The bins are the average values, and the error bars are the standard deviations of several samples. The twist angles are grouped by 5° for 0-5° and 55-60° and by 10° for 5-55°.

26